\documentclass[11pt,twoside]{article}
\usepackage{asp2010}

\resetcounters

\begin{document}

\title{Galactic foregrounds and CMB Polarization}
\author{Ettore Carretti$^1$
\affil{$^1$CSIRO Astronomy and Space Science, PO Box 276, Parkes, NSW 2870, Australia}
}

\begin{abstract}
The CMB polarization promises to unveil the dawn of time
measuring the gravitational wave background emitted 
by the Inflation. The CMB signal is faint, however, 
and easily contaminated by the Galactic foreground emission,
accurate measurements of which are thus crucial to make CMB
observations successful. We review the CMB
polarization properties and the current knowledge on the 
Galactic synchrotron emission, which dominates the foregrounds
budget at low frequency. We then focus on the S-Band Polarization 
All Sky Survey (S-PASS), a recently completed survey of the entire 
southern sky designed to investigate the Galactic CMB foreground.
\end{abstract}

\section{CMB polarization}
The Cosmic Microwave Background (CMB) is the radiation emitted
at the decoupling of matter and radiation as the temperature
of the Universe dropped below some 3000~K.  
Emitted in the early stages of  the Universe 
about 400 thousand years after the Big Bang, the CMB 
bears the signature of the primeval Universe conditions. 
The study of its temperature angular power spectrum carried out 
by a plethora of experiments (WMAP and the others, see e.g. \citealt{hinshaw09} 
and references therein) has been outstandingly 
successful measuring average properties of the Universe like geometry,
matter-energy content and nature, and expansion rate. It is worth noticing that 
the existence of the Dark Energy and its leading the matter-energy 
budget has been discovered thanks to the CMB anisotropies in combination
with high redshift Supernovae data.

\begin{figure}
\centering
  \includegraphics[angle=00, width=1.0\hsize]{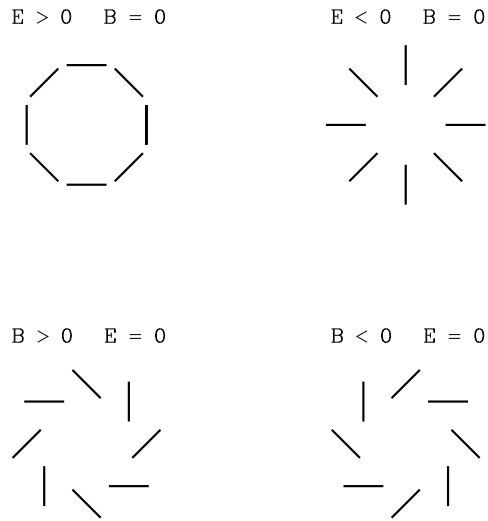}
\caption{Polarization angle patterns for pure $E$- (top) and $B$-Mode (bottom).
The former is characterised by a curl-free 
axisymmetric the polarization angle pattern, 
either radial or circular. A pure $B$-Mode, instead, is featured by polarization angles 
arranged in whirlpool-like structures either right- or left-handed.
\label{E-B-Mode:Fig}
}
\end{figure}

The study of the linearly polarized component is the next stage of CMB investigations. 
Usually described in terms of the 2-spin tensor Stokes parameters $Q$ and $U$, 
the polarized radiation is best studied in terms of $E$ and $B$-Modes in the case 
of the CMB~\citep[see][for a review]{zaldarriaga98}.  
These modes are defined by the pattern of the polarization angle field, which is axisymmetric for  
the $E$-Mode realising either radial or circular patterns, while is rotated by $45^\circ$ 
for the $B$-Mode realising whirlpool-like shapes (see Fig~\ref{E-B-Mode:Fig}). They 
have the undoubtable benefit to be scalar quantities.

\begin{figure}
\centering
  \includegraphics[angle=00, width=0.8\hsize]{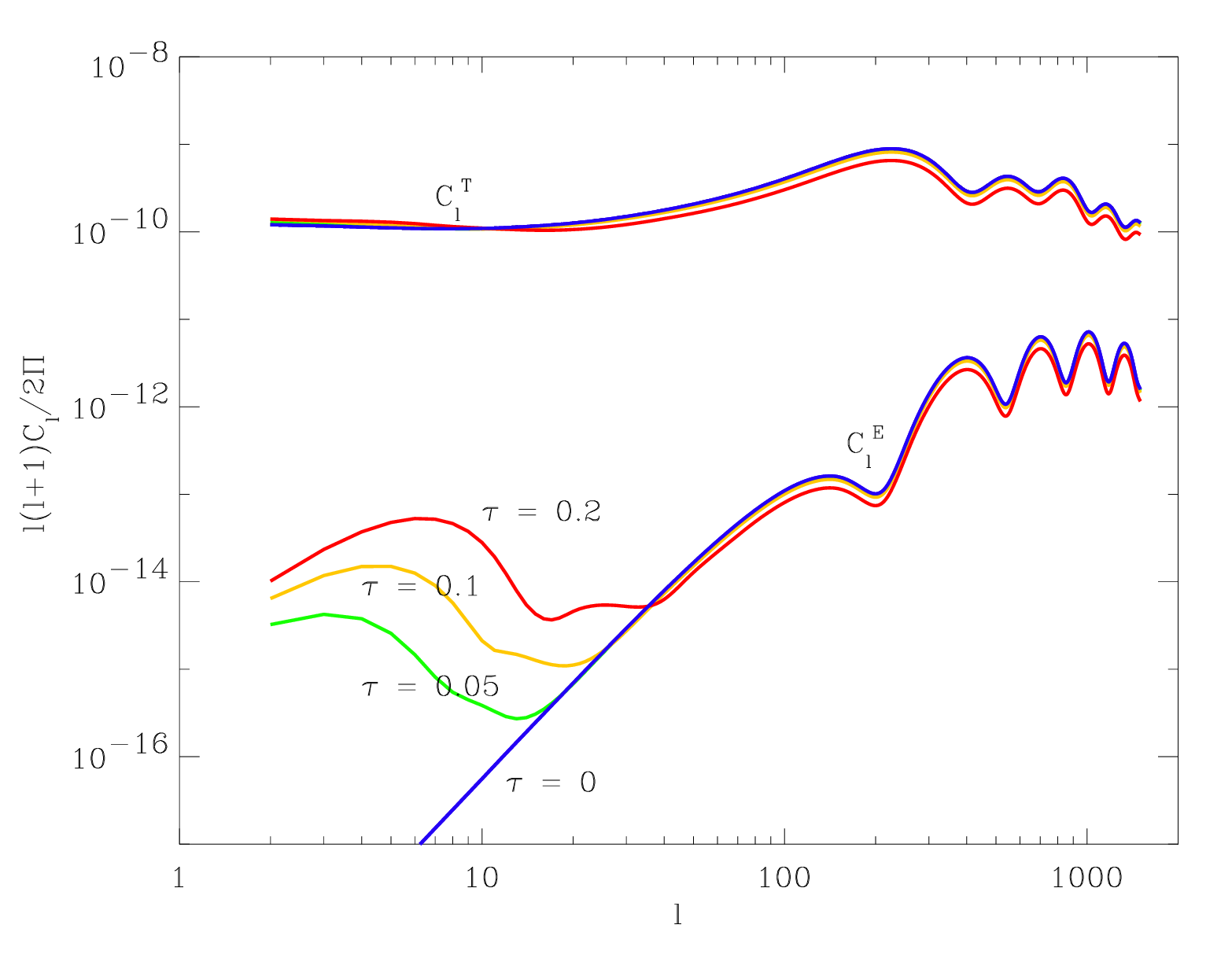}
\caption{CMB $E$-Mode power spectrum $C^E$ of three models differing for the 
optical depth $\tau$ of the CMB to the reionised medium. The spectrum is function
of the multipole $\ell$, related to the angular scale by $\theta \sim 180^\circ/\ell$. 
The amplitude and the position of the peak at the large angular 
scales ($\ell < 10$) measure $\tau$ and the redshift $z_{\rm ri}$ the reionisation
occurred at, respectively. The Temperature anisotropy spectra $C^T$ are also reported for
comparison.
\label{E-mode:Fig}
}
\end{figure}

The $E$-mode retains the signature of the reionisation history of the Universe. The 
large angular scales peak (Figure~\ref{E-mode:Fig}) is generated by the interaction 
of the CMB photons with the medium reionised at the end of the Dark Ages when
the first Galaxies formed and stars lit up.
The amplitude of the peak measures the optical depth of the medium to the CMB,
while its position tells about the redshift this event occurred at. 
That makes it a powerful tool to distinguish between several 
reionisation models.
\begin{figure}
\centering
  \includegraphics[angle=00, width=0.8\hsize]{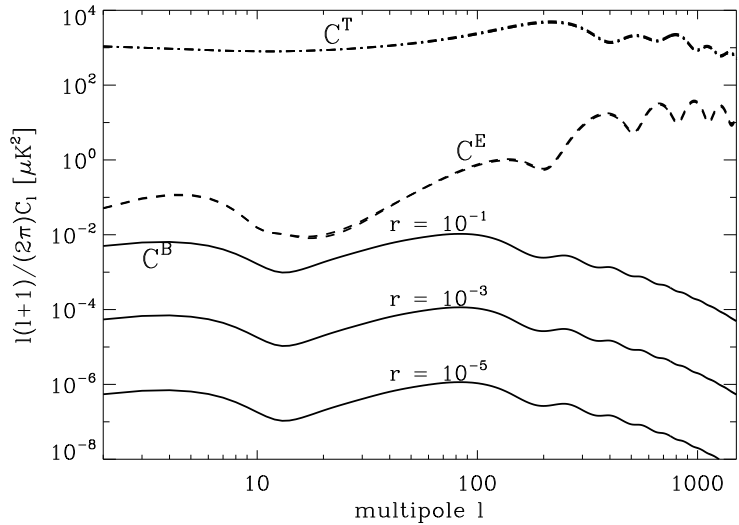}
\caption{CMB $B$-Mode power spectra for three models differing for the gravitational wave
power released by the Inflation. The amplitude is proportional to the gravitational wave power
measured by the tensor-to-scalar perturbation power ratio $r$ (see text). 
Temperature ($C^T$) and $E$-Mode spectra ($C^E$) are also shown.
\label{B-mode:Fig}
}
\end{figure}

But the most appealing feature of the CMB polarization is that the scalar (matter) 
perturbations  can generates only the $E$-mode component, so that
the $B$-Mode bears the tensor (gravitational waves) perturbations contribution 
uncontaminated by the much stronger scalar component. 
That way the CMB $B$--mode is a direct signature of the
primordial gravitational wave background (GWB) emitted by the Inflation
\citep[e.g.,][]{Kamionkowski98,boyle06}). 
The amplitude of its angular power spectrum is proportional to the GWB
power (Figure~\ref{B-mode:Fig}) and is usually measured in terms of  
fraction of the scalar perturbations by the tensor-to-scalar perturbation 
power ratio $r$. Still undetected, the current upper limit is set to 
$r < 0.20$ by the WMAP results~\citep[95\% C.L.,][]{komatsu09}.
Besides the evidence of a primordial GWB a measure of $r$ 
would help distinguish among several inflation models 
and investigate the physics of the very early stages 
of the Universe.

However, this spectacular science goal is made difficult 
by the CMB $B$-Mode tiny signal,
which ranges from about 100~nK of the current upper limit
down to 1~nK of the smallest $r$ accessible by CMB  \citep[$r\sim 2\times 10^{-5}$,][]{amarie05}.
Apart from the challenge of the required sensitivity, 
at such low levels the cosmic signal is easily overcome 
by the foreground Galactic synchrotron and dust emissions, 
which can set the actual detection limit.
The former leads the budget at low frequency and its study is essential 
to investigate  this cutting-edge fields of the current astrophysics research.

\section{Foregrounds: the Galactic Synchrotron emission}\label{synch:Sect}

\begin{figure}
\centering
  \includegraphics[angle=0, width=0.8\hsize]{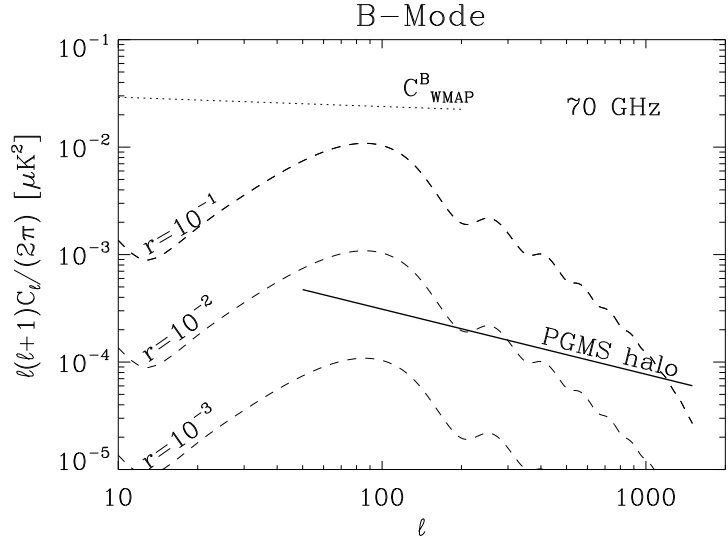}
\caption{Polarized Galactic emission at high Galactic latitudes at 70~GHz:
               average of the entire high Galactic latitudes as estimated by WMAP 23~GHz data 
               ($C^B_{\rm WMAP}$, dotted)
               and in the best 15\% of the sky as estimated in the halo section of the PGMS survey  (solid).
               CMB $B$--Mode spectra for different $r$ values are also reported for comparison (dashed).
                \label{status:Fig}
}
\end{figure}

Until a few years ago the Galactic synchrotron emission at 
high Galactic latitude was mostly unknown with large uncertainties even 
about its emission level, making unclear the actual limit to detect the CMB polarized component.
A number of observations have been conducted since early 2000s to fill this gap 
and now the overall picture seems to have been understood (Figure~\ref{status:Fig}). 

WMAP data have allowed for the first time to study the average emission at 
high Galactic latitude, the region useful for CMB observation because of the
lower Galactic emission. 
\citet{page07} have analysed the 23~GHz WMAP polarized maps and
find that the normal emission at high Galactic latitude is strong: scaled to 70~GHz
 it is {\it equivalent to}\footnote{With \"{\it equivalent to $r$}" we intend the 
                                               value of $r$ for which the CMB $B$--Mode spectrum at the $\ell = 90$ peak 
                                               matches that of the foreground.}
$r \sim 0.3$, which is even higher than the current upper limit (Figure~\ref{status:Fig}). 
A similar result has been obtained by \citet{laporta06}, who, 
using the 1.4~GHz all-sky map combination of the 
DRAO and Villa Elisa surveys \citep{wolleben06,testori08}, 
estimate an emission at 70~GHz equivalent to $r = 0.5$.

Better conditions have been found in the lowest emission regions of the sky. 
A first  analysis to understand position and extension of the 
best regions has been carried out by \citet{carretti06} with WMAP data 
 and shows that those regions cover a non-negligible 
part of the sky (about 15\%). WMAP data have not been sufficient sensitivity 
to study the properties of the signal of these areas. A detailed analysis 
has been carried out only recently with the Parkes Galactic Meridian Survey 
(PGMS, \cite{carretti10_pgms}),  a survey of a 5-degree wide 
strip stretching from the Galactic plane to the south Galactic pole  
 through the southern portion of these lowest emission areas.
  
\begin{figure}
\centering
  \includegraphics[angle=0, width=0.8\hsize]{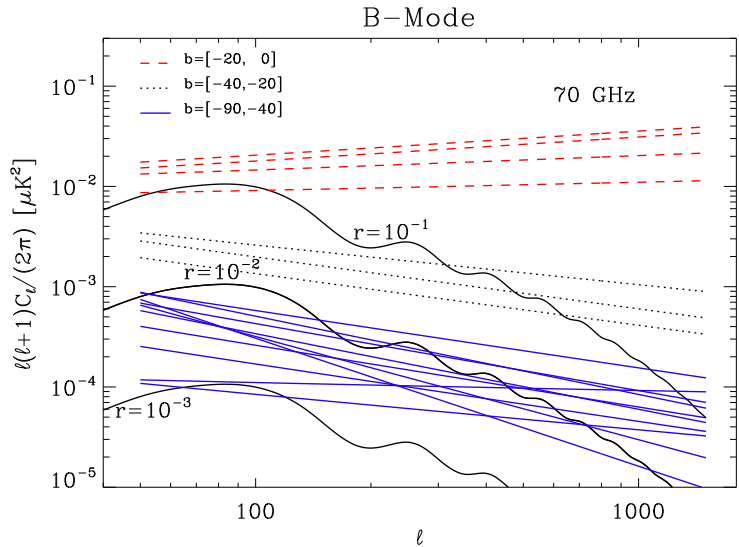}
\caption{Polarized angular power spectra of the 17~PGMS fields grouped by 
                their Galactic latitude $b$ \citep{carretti10_pgms}.
                The spectra are scaled to 70~GHz to estimate their level in the CMB frequency 
                window. CMB $B$-Mode spectra for three values of $r$ are also reported for comparison 
                (solid black).
\label{pgms:Fig}
}
\end{figure}

 Conducted at 2.3~GHz with the Parkes telescope, it was aimed at studying the 
 polarized emission behaviour versus the Galactic latitude
 in a region uncontaminated by large local structures like the big radio loops.
The PGMS has identified three latitude sections distinguished: 
the disc, the halo, and a transition region connecting them (Figure~\ref{pgms:Fig}). 
The halo section lies at latitudes $|b| > 40^\circ$ and has weak and smooth polarized
emission mostly at large scale with steep angular power spectra, 
while the disc region covers the latitudes $|b|<20\circ$ and has a brighter,
more complex emission dominated by the small scales with flatter spectra. 
The transition region has steep spectra as in the halo, but the emission 
increases toward the Galactic plane from halo to disc levels.  
The most interesting result is that the halo section does not show 
an obvious trend with respect to the Galactic latitude resulting in just one 
environment stretching over $50\circ$ with very low emission. 
This, once scaled to 70~GHz, is equivalent to the CMB $B$--Mode emission
of models with $r_{\rm halo} = (3.3\pm0.4)\times10^{-3}$ (Figure~\ref{status:Fig}). This
are very good conditions for the detection of the $B$-mode both in terms 
of emission level and size of the area.

Therefore, the current picture shows a strong synchrotron contamination
at high Galactic latitudes with a typical emission even higher than the
current upper limit of the CMB $B$--Mode. This requires an aggressive 
foreground cleaning to search for the inflation signature with all-sky class surveys. 
Much better conditions exist in the best 15\% of the sky, where the detection limit
of $r$ drops to $\delta r = 2\times 10^{-3}$ (3-$\sigma$ C.L.).  This 
matches the needs of the forthcoming generation of ground-based and 
balloon-borne experiments ($r \sim 0.01$, e.g.: SPIDER, EBEX, 
and QUIET, \cite{crill08,grainger08,samtleben08}) and approaches 
the detection limits set for the next generation space missions 
currently under study such as B-POL and CMBPol \citep{debernardis09,baumann08}.

Because of its implications for the experiment design another important point to account 
for is the frequency of minimum foreground emission where synchrotron and dust 
equate. 
WMAP results show that this is in the range 60--70~GHz on average at 
high Galactic latitudes \cite{page07}. The PGMS analysis shows that it 
should set in a similar range also in the weakest emission areas (60--80~GHz).
If confirmed in other regions, this would imply that dust and synchrotron emission 
are mostly correlated across the sky and that  the
frequency of minimum foreground is nearly independent of the sky position. 
This is still an open point, however, because the polarized dust emission has not 
yet been detected in the PGMS area (or in other low emission areas) 
and the results are based on assumptions about the dust polarization fraction.

\section{Future Perspectives}

The high level of contamination at high Galactic latitude (even higher than the 
current CMB upper limit) requires aggressive cleaning and, in turn, sensitive 
and accurate foreground maps in case all-sky class CMB mission 
want to be carried out.
The identification of the best regions of the sky 
and a full characterisation of their emission properties 
also call for sensitive all-sky class foreground surveys. 
The latter is essential for sub-orbital experiments (ground-based 
and balloon-borne), which target their observations to small areas 
with low foreground emission. 

\begin{figure}
\centering
  \includegraphics[angle=90, width=0.8\hsize]{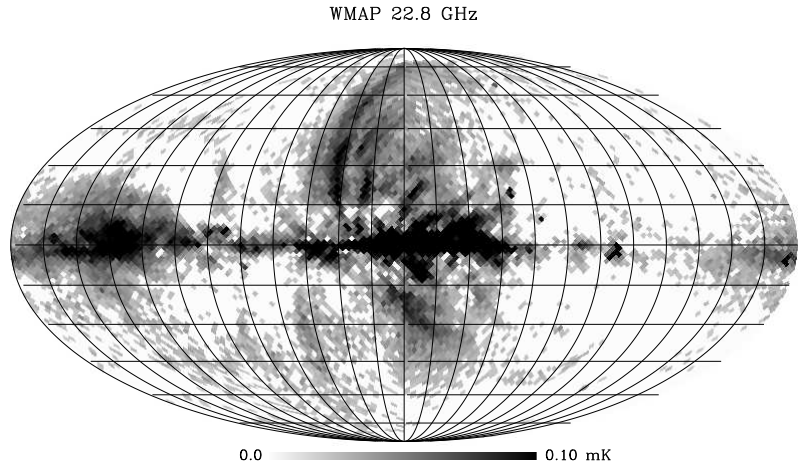}
\caption{WMAP polarized intensity map at 23~GHz. The data have been binned in $2^\circ$
               pixels and all the pixels with $S/N < 3$ have been blanked (white).
               The map is in Galactic coordinates centred at the Galactic Centre.
                \label{wmap:Fig}
}
\end{figure}

Even though the investigations conducted so far have clarified the picture 
of the contamination by synchrotron emission, the currently available data 
set is not sufficient for such jobs, however. 
As mentioned in Sect.~\ref{synch:Sect}, the 23~GHz map by WMAP 
has allowed a statistical detection at high Galactic latitude, 
but even smoothing the data on the 2~degree scale of the $B$-Mode peak the 
signal-to-noise ratio is $S/N < 3$ in about 55\% of the sky. 
As shown by Figure~\ref{wmap:Fig}, this is mostly 
the entire area useful for CMB observations and consists of
all the high Galactic latitudes except the areas 
contaminated by large local structures like the big radio loops. 

\begin{figure}
\centering
  \includegraphics[angle=90, width=0.8\hsize]{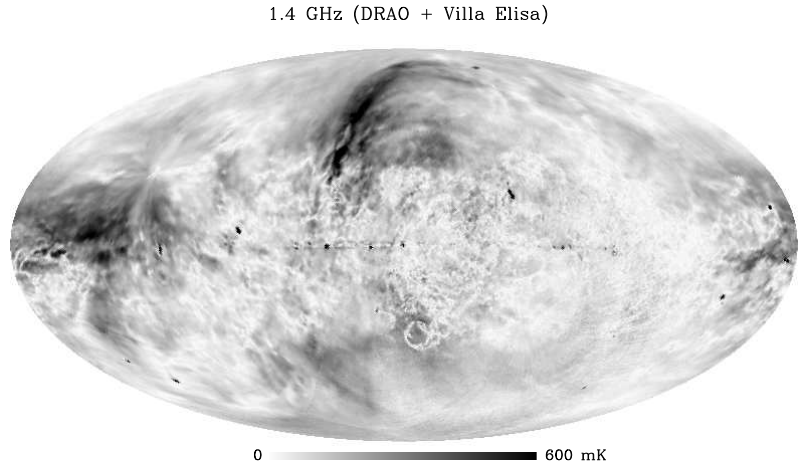}
\caption{Polarized intensity map at 1.4 GHz obtained combining the DRAO and Villa Elisa surveys.
                \label{drao:Fig}
}
\end{figure}

The 1.4~GHz all-sky map combination of the DRAO and Villa Elisa surveys (Figure~\ref{drao:Fig})
benefits of a better $S/N$, but modifications by Faraday Rotation (FR) affect the data. 
All the Galactic disc at latitude $|b|<30^\circ$ is strongly depolarized and Faraday 
modulation is present up to $|b| = 50^\circ$ (\cite{carretti05}),
making this data set not suitable to be extrapolated to the CMB frequency window and 
used as template for foreground separation.

The PGMS has the right sensitivity for cleaning purposes ($S/N>7$ on 9~arcmin 
scale even in the lowest emission areas observed), but it has surveyed only a 
strip and  is not sufficient to give a template map of the entire low emission regions. 

More data are thus required. The observations conducted so far lead to the following 
main requirements  for new surveys:
\begin{enumerate}  
  \item{} to be conducted at a low enough radio frequency for the 
               synchrotron emission to dominate
               the other diffuse emission components;
 \item{} at a frequency higher than 1.4~GHz to avoid
               significant Faraday Rotation effects;
 \item{}  at a frequency not too high not to compromise the $S/N$ ratio;
 \item{} to be an all-sky class survey to be used as template for CMB space 
            missions or to identify and characterise the areas with lowest Galactic
             emission;
\end{enumerate}  

There are two major surveys just completed or are currently on-going 
that satisfy such specifications: the S-band Polarization All Sky Survey (S-PASS) and 
the C-Band All-Sky Survey (C-BASS). 

S-PASS is a survey of the entire southern sky at 2.3~GHz 
conducted with the Parkes telescope (see Sect.~\ref{spass:Sect}). 
It has an angular resolution of 9~arcmin which covers all the angular 
scale range required by CMB observations. 
It has has been recently completed and features a high 
$S/N$ ratio that can enable the cleaning procedures to 
extend the $B$-mode detection limit down to the lowest 
value of $r$ accessible by CMB.
  
C-BASS~\citep{muchovej10} covers the entire sky and it is complementary to 
S-PASS. It is conducted at higher frequency 
(5~GHz), but sports a lower S/N ratio (5~times lower by specifications) 
and a coarser resolution (about 1~degree), which enables it to cover the 
$2^\circ$ CMB peak but does not stretch down to the 10~arcmin required by 
the gravitational lensing component. The observations are planned to start
in 2010 and will last for a couple of years.

The combination of these two surveys promises to give a solid data base for 
the aggressive cleaning required to cope with the strong Galactic signal. S-PASS
gives a sounding ground thanks to its high $S/N$, while the combination with C-BASS 
ensures the determination of the accurate frequency spectral index 
required for safe extrapolations to the CMB frequency window.

In the next Sections we will discuss about S-PASS, which has been recently 
completed and has already provided preliminary maps.

\section{S-PASS}~\label{spass:Sect}

\subsection{The survey}
The S-band Polarization All Sky Survey (S-PASS) is a project mapping the diffuse
polarized synchrotron emission of the entire southern sky with the Parkes
radio telescope at 2.3~GHz \citep{carretti07}. Commenced in
October 2007, the observations has been completed in January 2010 and 
the data reduction is in progress. Besides studying the CMB foregrounds 
the survey is aimed at investigating the Galactic magnetic 
field in the halo, in the disc, and at the disc-halo transition.

The resolution of 9~arcmin makes S-PASS ideal for CMB foregrounds
 investigations covering the entire angular scale range required. The sensitivity 
of 1~mK per beam-sized pixel gives $S/N >3$ everywhere, and, for the sake of 
comparison with the existing data, corresponds to  $S/N > 20$ on 1-deg scale. 
With such performances, S-PASS can clean CMB maps 
enabling $r$ detection limits of $\delta r = 4\times 10^{-5}$ 
at 3-$\sigma$ C.L. \citep[we use the estimators of][]{tucci05}.
This is close to the lowest detection limit possible by the CMB $B$-Mode, 
so that S-PASS is a sort of ultimate synchrotron foreground survey 
in terms of sensitivity requirements.
  
A survey with such characteristics has posed several challenges to 
make it feasible with a large telescope like the 64-m dish of Parkes. 
For instance, for the scientific goals it is essential to detect the signal also in the 
lowest emission regions, where values of a few mK were expected. This makes the 
ground emission contamination a serious issue, especially for the need to 
preserve the signal up to the largest angular scales for absolute calibration purposes. 
Another example is the angular resolution, which with its 9~arcmin is a factor four 
better than that of the 1.4~GHz survey. This gives far more details, but, at the same time, 
requires four times more scans and observing time to fully sample the sky if conducted 
with a standard observing mode. Moreover, the Galactic physics goals 
set as key requirement for S-PASS the absolute calibration, a long standing 
issue of radioastronomical observations both because of instrumental offsets 
and ground emission contamination. 

\subsection{Scanning strategy and absolute calibration}

To cope with these challenges a new non-standard observing strategy 
based on long and fast azimuth scans has been developed. 
A detailed description will be presented in a forthcoming 
paper \citep{carretti10_spass}, here we briefly report the key futures.

The azimuth scans have been chosen to minimise ground emission 
pick-up variations. In S-PASS data these are of the order of a few tens mK, 
against the typical variations of a few hundreds mK of previous surveys 
\citep[e.g., cfr.][]{wolleben06}, which makes obvious the benefits of azimuth 
scans.

The scan length (about $115^\circ$) and elevation (Celestial south pole at Parkes) allow 
covering the entire DEC range in just one scan preserving the information on all the angular scales. 

This type of scan has also required to set up a special observing mode based on an uninterrupted 
sequence of back and forth scans with both no loss of tracking at the turnoff and precise timing 
to enable a regular gridding of the sky. 
The Earth rotation lets the sky to drift in RA so that the entire 24-hour range can  be covered in
one sidereal day. As a result, a zig-zag in the sky is observed each night. The 
combination of  observations taken in a number of nights appropriately offset in RA 
allows covering the entire southern sky.

The vast area to observe and the small size of the pixels (4.5') has required
a high scan speed to conduct the survey in an acceptable time duration. 
The scan rate has been pushed up to 15$^\circ$/min, 
which is a significant fraction of the slewing speed (24$^\circ$/min), not a bad performance for
a 1000-tons 64-m telescope.  The use of long scans has allowed minimising the overhead to 
ramp up and down the telescope speed at the two scan ends, which is quite significant
when using high scan rates. This, in combination with the high speed, has allowed
to complete the survey in less than 2000-h of telescope time.

\begin{figure}
\centering
  \includegraphics[angle=90, width=0.9\hsize]{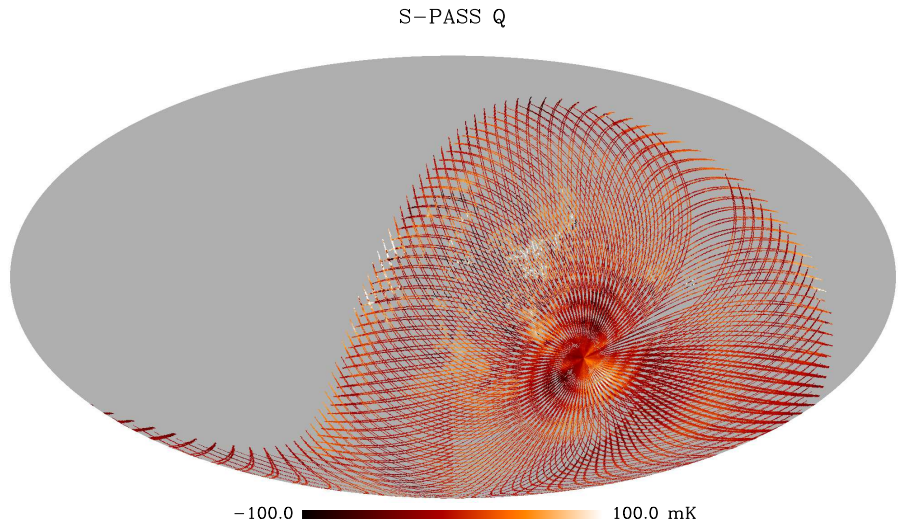}
\caption{Combination of two set of scans taken observing eastward and westward 
               respectively. The crossing between scans of the two sets is obvious and
               enables effective basket-weaving.
                \label{basket:Fig}
}
\end{figure}

The data reduction of radioastronomical data usually requires baseline subtraction to
remove both ground emission and instrumental offsets, which also removes 
the mean level of the signal over the scan. To recover the power on all angular scales 
a "basket weaving"  technique (scan crossing) is required.
Basket weaving is a bit tricky with AZ scans. However, the same DEC range 
can be observed both
when the sky rises and sets. We thus observed two sets of scans,
one eastward and the other westward, whose combination realises 
an effective scan crossing (see Figure~\ref{basket:Fig}).
\begin{figure}
\centering
  \includegraphics[angle=270, width=0.9\hsize]{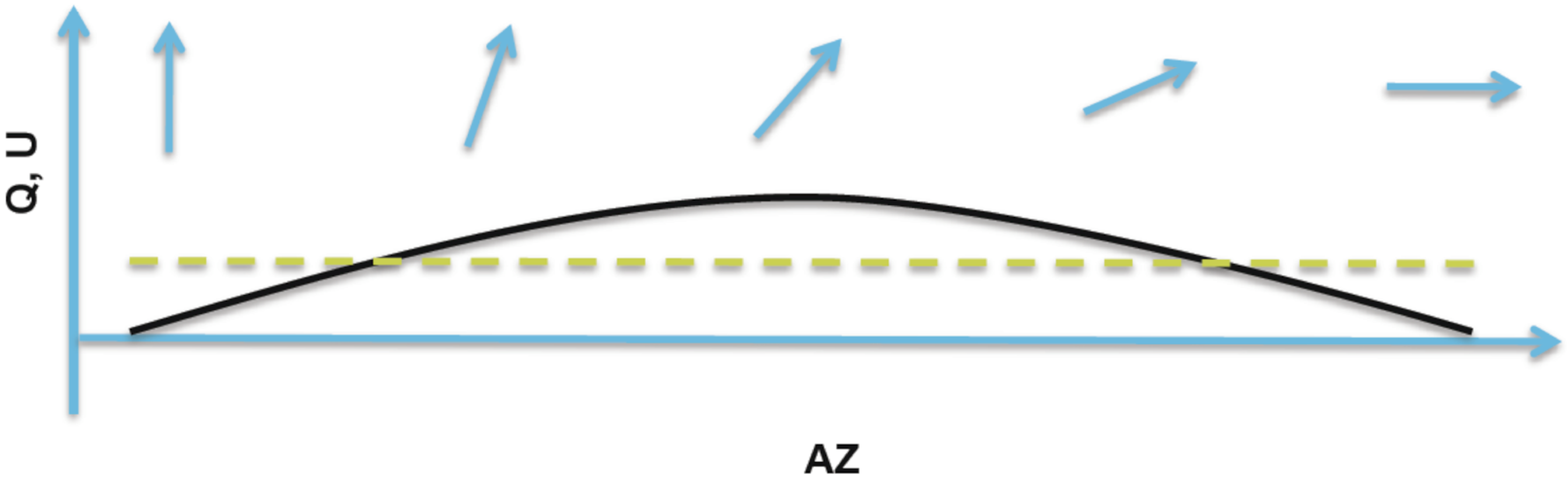}
\caption{The parallactic angle variation along a long AZ scan (arrows)
                modulates a signal constant in the sky reference frame
                into a section of sinusoid in the instrument reference frame (solid) so that
                the baseline subtraction (dashed) does not completely remove it.
                Very long scans are essential.
                \label{absolute:Fig}
}
\end{figure}

However, this is not yet sufficient to absolute calibrate the data since 
the basket-weaving can recover the power up to the scale 
of the map-size, but leaving the average signal still undetermined.

The new idea developed for S-PASS to solve this point has been to use the 
parallactic angle modulation of Stokes $Q$ and $U$ along a scan. 
Usually, a constant signal would be removed by the baseline subtraction, 
but the change of parallactic angle along a very long scan modulates 
both Q and U as sections of sinusoid in the instrument reference frame
(Figure~\ref{absolute:Fig}). 
That way, the baseline subtraction preserves most of the Q and U average 
signal. An appropriate inverse problem solving is able to recover
the average signal and then absolutely calibrate the
map (see \citep{carretti10_spass} for full details and tests).
Simulations conducted using the actual sensitivity, 
signal, and scanning strategy of S-PASS show that the absolute level 
can be reconstructed with a precision of about 50~$\mu$K, far better than
the survey statistical noise, thus giving a negligible contribution 
to the overall error budget.
It is worth noticing that conducting uninterrupted very long scans 
is essential, otherwise the parallactic angle modulation would be 
marginal and the method ineffective.

\subsection{Preliminary maps}

\begin{figure}
\centering
  \includegraphics[angle=90, width=0.9\hsize]{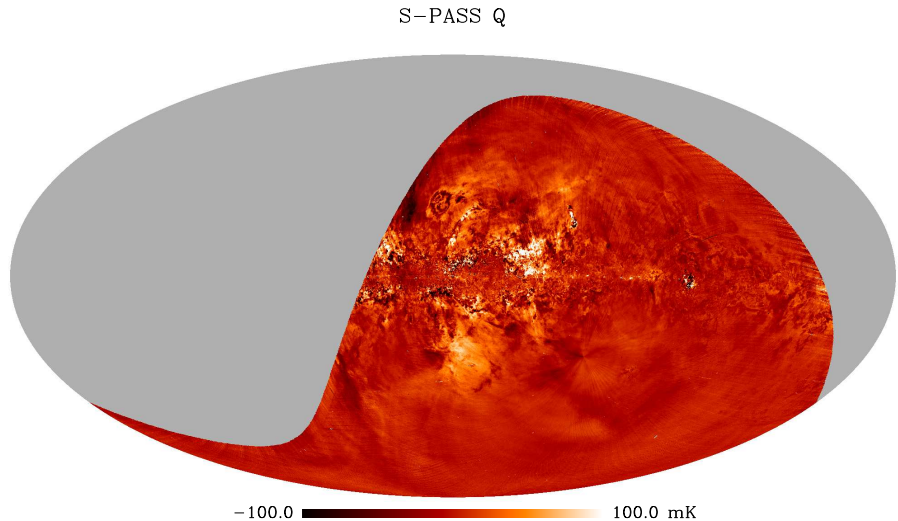}
\caption{S-PASS Stokes $Q$ map. The full map-making procedure is not yet applied 
                here.
\label{spass:Fig}
}
\end{figure}

Figure~\ref{spass:Fig} shows a preliminary map of Stokes $Q$ at full resolution 
obtained using the entire data but without applying the full 
map-making procedure to recover the absolute calibration. (That way, stripes are still visible.)
The map shows plenty of details and structures unseen 
at lower frequencies. The emission is smooth not only at high latitudes but also
in the disc down to $|b| = 5^\circ$--$10^\circ$ even in the inner Galaxy. This
will enable first studies of the disc and the disc-halo transition with diffuse emission
data.

The absolutely calibrated maps (not shown here) are featured by very large scale
structures up to the size of the map which matches those of the WMAP 23~GHz map 
both in shape and position. That is both evidence of the quality of the overall 
data reduction procedure and that Faraday Rotation effects are marginal in the halo 
already at this frequency. The Faraday modulation 
 would have either modified the structure shape or shifted their positions
because of polarization angle shift. Therefore, the S-PASS data 
can be already used for CMB foreground cleaning purposes.

A number of features are visible on the Galactic plane too. 
At both sides of the Galactic Centre there are two large 
areas fully depolarized corresponding to two large local HII emission regions 
visible in the SHASSA map \citep{gaustad01}. Their view at full resolution is dramatic: 
the signal is almost absent because fully depolarized, but it reappears 
at their edges as a narrow ring of strongly modulated emission, evidence that the FR
is lower and cannot fully depolarize the signal but can still modulate it strongly. 
After that transition the signal is smooth, meaning that the FR effects are much lower. 
Always in the Galactic plane but In the outer Galaxy is visible a intriguing 
mottled region at $l = [230^\circ, 280^\circ]$ and $|b| < 15^\circ$ 
in correspondence of the Gum Nebula. Other objects like 
the Vela SNR, Centaurs A, Large Magellanic Cloud, and Fornax A are
obvious. 

In summary, S-PASS is revealing a new polarized sky. 
The disc emission looks unveiled at last, making possible investigations 
of this part of the Galaxy, while the high S/N at high latitude promises both a leap 
in the foreground cleaning efficiency and a better understanding 
of the large scale Galactic magnetic field in the halo. 

\acknowledgements{We thanks the organisers for the excellent 
                                      conference which has offered a stimulating
                                      environment for many discussions on ISM topics.
                                      Part of this work is based on data taken with the Parkes telescope.
                                      The Parkes telescope is part of the Australia Telescope which 
                                      is funded by the Commonwealth of Australia for operation as 
                                      a National Facility managed by CSIRO.
                                      We acknowledge the use of the the packages CMBFAST and
                                      HEALPix~\citep{gorski05} and the use of the Legacy Archive for Microwave Background Data Analysis 
                                      (LAMBDA). Support for LAMBDA is provided by the NASA Office of Space Science.}
                                      
\bibliography{carretti_ettore}

\begin{thebibliography}{}
\expandafter\ifx\csname natexlab\endcsname\relax\def\natexlab#1{#1}\fi
\expandafter\ifx\csname url\endcsname\relax
  \def\url#1{\texttt{#1}}\fi
\expandafter\ifx\csname urlprefix\endcsname\relax\def\urlprefix{URL }\fi
\providecommand{\eprint}[2][]{\url{#2}}

\bibitem[{{Amarie} et~al.(2005){Amarie}, {Hirata}, \& {Seljak}}]{amarie05}
{Amarie}, M., {Hirata}, C., \& {Seljak}, U. 2005, \prd, 72, 123006.
  \eprint{arXiv:astro-ph/0508293}

\bibitem[{{Baumann} et~al.(2008){Baumann}, {Cooray}, {Dodelson}, {Dunkley},
  {Fraisse}, {Jackson}, {Kogut}, {Krauss}, {Smith}, \&
  {Zaldarriaga}}]{baumann08}
{Baumann}, D., {Cooray}, A., {Dodelson}, S., {Dunkley}, J., {Fraisse}, A.~A.,
  {Jackson}, M.~G., {Kogut}, A., {Krauss}, L.~M., {Smith}, K.~M., \&
  {Zaldarriaga}, M. 2008, arXiv:0811.3911

\bibitem[{{Boyle} et~al.(2006){Boyle}, {Steinhardt}, \& {Turok}}]{boyle06}
{Boyle}, L.~A., {Steinhardt}, P.~J., \& {Turok}, N. 2006, Physical Review
  Letters, 96, 111301. \eprint{arXiv:astro-ph/0507455}

\bibitem[{{Carretti} et~al.(2006){Carretti}, {Bernardi}, \&
  {Cortiglioni}}]{carretti06}
{Carretti}, E., {Bernardi}, G., \& {Cortiglioni}, S. 2006, \mnras, 373, L93.
  \eprint{arXiv:astro-ph/0609288}

\bibitem[{{Carretti} et~al.(2005){Carretti}, {Bernardi}, {Sault},
  {Cortiglioni}, \& {Poppi}}]{carretti05}
{Carretti}, E., {Bernardi}, G., {Sault}, R.~J., {Cortiglioni}, S., \& {Poppi},
  S. 2005, \mnras, 358, 1. \eprint{arXiv:astro-ph/0412598}

\bibitem[{{Carretti} et~al.(2010{\natexlab{a}}){Carretti}, {Haverkorn},
  {McConnell}, {Bernardi}, {McClure-Griffiths}, {Cortiglioni}, \&
  {Poppi}}]{carretti10_pgms}
{Carretti}, E., {Haverkorn}, M., {McConnell}, D., {Bernardi}, G.,
  {McClure-Griffiths}, N.~M., {Cortiglioni}, S., \& {Poppi}, S.
  2010{\natexlab{a}}, \mnras, 405, 1670. \eprint{0907.4861}

\bibitem[{{Carretti} et~al.(2010{\natexlab{b}}){Carretti}, {Kesteven},
  {Bernardi}, {Cortiglioni}, {Gaensler}, {Haverkorn}, {Poppi}, \&
  {Staveley-Smith}}]{carretti10_spass}
{Carretti}, E., {Kesteven}, M., {Bernardi}, G., {Cortiglioni}, S., {Gaensler},
  B., {Haverkorn}, M., {Poppi}, S., \& {Staveley-Smith}, L. 2010{\natexlab{b}},
  in preparation

\bibitem[{{Carretti} et~al.(2007){Carretti}, {Staveley-Smith}, {Haverkorn},
  {Bernardi}, {Cortiglioni}, {Gaensler}, {Kesteven}, \& {Poppi}}]{carretti07}
{Carretti}, E., {Staveley-Smith}, L., {Haverkorn}, M., {Bernardi}, G.,
  {Cortiglioni}, S., {Gaensler}, B., {Kesteven}, M., \& {Poppi}, S. 2007,
  Parkes Telescope Project, P560

\bibitem[{{Crill} et~al.(2008){Crill}, {Ade}, {Battistelli}, {Benton},
  {Bihary}, {Bock}, {Bond}, {Brevik}, {Bryan}, {Contaldi}, {Dor{\'e}},
  {Farhang}, {Fissel}, {Golwala}, {Halpern}, {Hilton}, {Holmes}, {Hristov},
  {Irwin}, {Jones}, {Kuo}, {Lange}, {Lawrie}, {MacTavish}, {Martin}, {Mason},
  {Montroy}, {Netterfield}, {Pascale}, {Riley}, {Ruhl}, {Runyan}, {Trangsrud},
  {Tucker}, {Turner}, {Viero}, \& {Wiebe}}]{crill08}
{Crill}, B.~P., {Ade}, P.~A.~R., {Battistelli}, E.~S., {Benton}, S., {Bihary},
  R., {Bock}, J.~J., {Bond}, J.~R., {Brevik}, J., {Bryan}, S., {Contaldi},
  C.~R., {Dor{\'e}}, O., {Farhang}, M., {Fissel}, L., {Golwala}, S.~R.,
  {Halpern}, M., {Hilton}, G., {Holmes}, W., {Hristov}, V.~V., {Irwin}, K.,
  {Jones}, W.~C., {Kuo}, C.~L., {Lange}, A.~E., {Lawrie}, C., {MacTavish},
  C.~J., {Martin}, T.~G., {Mason}, P., {Montroy}, T.~E., {Netterfield}, C.~B.,
  {Pascale}, E., {Riley}, D., {Ruhl}, J.~E., {Runyan}, M.~C., {Trangsrud}, A.,
  {Tucker}, C., {Turner}, A., {Viero}, M., \& {Wiebe}, D. 2008, in Society of
  Photo-Optical Instrumentation Engineers (SPIE) Conference Series, vol. 7010
  of Society of Photo-Optical Instrumentation Engineers (SPIE) Conference
  Series. \eprint{0807.1548}

\bibitem[{{De Bernardis} et~al.(2009){De Bernardis}, {Bucher}, {Burigana}, \&
  {Piccirillo}}]{debernardis09}
{De Bernardis}, P., {Bucher}, M., {Burigana}, C., \& {Piccirillo}, L. 2009,
  Experimental Astronomy, 23, 5. \eprint{0808.1881}

\bibitem[{{Gaustad} et~al.(2001){Gaustad}, {McCullough}, {Rosing}, \& {Van
  Buren}}]{gaustad01}
{Gaustad}, J.~E., {McCullough}, P.~R., {Rosing}, W., \& {Van Buren}, D. 2001,
  \pasp, 113, 1326. \eprint{arXiv:astro-ph/0108518}

\bibitem[{{G{\'o}rski} et~al.(2005){G{\'o}rski}, {Hivon}, {Banday}, {Wandelt},
  {Hansen}, {Reinecke}, \& {Bartelmann}}]{gorski05}
{G{\'o}rski}, K.~M., {Hivon}, E., {Banday}, A.~J., {Wandelt}, B.~D., {Hansen},
  F.~K., {Reinecke}, M., \& {Bartelmann}, M. 2005, \apj, 622, 759.
  \eprint{arXiv:astro-ph/0409513}

\bibitem[{{Grainger} et~al.(2008){Grainger}, {Aboobaker}, {Ade}, {Aubin},
  {Baccigalupi}, {Bissonnette}, {Borrill}, {Dobbs}, {Hanany}, {Hogen-Chin},
  {Hubmayr}, {Jaffe}, {Johnson}, {Jones}, {Klein}, {Korotkov}, {Leach}, {Lee},
  {Levinson}, {Limon}, {Macaluso}, {MacDermid}, {Matsumura}, {Meng}, {Miller},
  {Milligan}, {Pascale}, {Polsgrove}, {Ponthieu}, {Reichborn-Kjennerud},
  {Renbarger}, {Sagiv}, {Stivoli}, {Stompor}, {Tran}, {Tucker}, {Vinokurov},
  {Zaldarriaga}, \& {Zilic}}]{grainger08}
{Grainger}, W., {Aboobaker}, A.~M., {Ade}, P., {Aubin}, F., {Baccigalupi}, C.,
  {Bissonnette}, {\'E}., {Borrill}, J., {Dobbs}, M., {Hanany}, S.,
  {Hogen-Chin}, C., {Hubmayr}, J., {Jaffe}, A., {Johnson}, B., {Jones}, T.,
  {Klein}, J., {Korotkov}, A., {Leach}, S., {Lee}, A., {Levinson}, L., {Limon},
  M., {Macaluso}, J., {MacDermid}, K., {Matsumura}, T., {Meng}, X., {Miller},
  A., {Milligan}, M., {Pascale}, E., {Polsgrove}, D., {Ponthieu}, N.,
  {Reichborn-Kjennerud}, B., {Renbarger}, T., {Sagiv}, I., {Stivoli}, F.,
  {Stompor}, R., {Tran}, H., {Tucker}, G., {Vinokurov}, J., {Zaldarriaga}, M.,
  \& {Zilic}, K. 2008, in Society of Photo-Optical Instrumentation Engineers
  (SPIE) Conference Series, vol. 7020 of Society of Photo-Optical
  Instrumentation Engineers (SPIE) Conference Series

\bibitem[{{Hinshaw} et~al.(2009){Hinshaw}, {Weiland}, {Hill}, {Odegard},
  {Larson}, {Bennett}, {Dunkley}, {Gold}, {Greason}, {Jarosik}, {Komatsu},
  {Nolta}, {Page}, {Spergel}, {Wollack}, {Halpern}, {Kogut}, {Limon}, {Meyer},
  {Tucker}, \& {Wright}}]{hinshaw09}
{Hinshaw}, G., {Weiland}, J.~L., {Hill}, R.~S., {Odegard}, N., {Larson}, D.,
  {Bennett}, C.~L., {Dunkley}, J., {Gold}, B., {Greason}, M.~R., {Jarosik}, N.,
  {Komatsu}, E., {Nolta}, M.~R., {Page}, L., {Spergel}, D.~N., {Wollack}, E.,
  {Halpern}, M., {Kogut}, A., {Limon}, M., {Meyer}, S.~S., {Tucker}, G.~S., \&
  {Wright}, E.~L. 2009, \apjs, 180, 225. \eprint{0803.0732}

\bibitem[{{Kamionkowski} \& {Kosowsky}(1998)}]{Kamionkowski98}
{Kamionkowski}, M., \& {Kosowsky}, A. 1998, \prd, 57, 685.
  \eprint{arXiv:astro-ph/9705219}

\bibitem[{{Komatsu} et~al.(2009){Komatsu}, {Dunkley}, {Nolta}, {Bennett},
  {Gold}, {Hinshaw}, {Jarosik}, {Larson}, {Limon}, {Page}, {Spergel},
  {Halpern}, {Hill}, {Kogut}, {Meyer}, {Tucker}, {Weiland}, {Wollack}, \&
  {Wright}}]{komatsu09}
{Komatsu}, E., {Dunkley}, J., {Nolta}, M.~R., {Bennett}, C.~L., {Gold}, B.,
  {Hinshaw}, G., {Jarosik}, N., {Larson}, D., {Limon}, M., {Page}, L.,
  {Spergel}, D.~N., {Halpern}, M., {Hill}, R.~S., {Kogut}, A., {Meyer}, S.~S.,
  {Tucker}, G.~S., {Weiland}, J.~L., {Wollack}, E., \& {Wright}, E.~L. 2009,
  \apjs, 180, 330. \eprint{0803.0547}

\bibitem[{{La Porta} et~al.(2006){La Porta}, {Burigana}, {Reich}, \&
  {Reich}}]{laporta06}
{La Porta}, L., {Burigana}, C., {Reich}, W., \& {Reich}, P. 2006, \aap, 455,
  L9. \eprint{arXiv:astro-ph/0607300}

\bibitem[{{Muchovej} et~al.(2010){Muchovej}, {All Sky Survey}, {Pearson},
  {Stevenson}, {Readhead}, {Leitch}, {Jones}, {Lawrence}, {Rocha}, {King},
  {Taylor}, {Jones}, {Holler}, {Davis}, {Dickinson}, {Jaffe}, {Leahy},
  {Copley}, {Jonas}, {Booth}, {Hafez}, \& {Almeqren}}]{muchovej10}
{Muchovej}, S., {All Sky Survey}, C., {Pearson}, T., {Stevenson}, M.,
  {Readhead}, T., {Leitch}, E., {Jones}, D., {Lawrence}, C., {Rocha}, G.,
  {King}, O., {Taylor}, A., {Jones}, M., {Holler}, C., {Davis}, R.,
  {Dickinson}, C., {Jaffe}, T., {Leahy}, P., {Copley}, C., {Jonas}, J.,
  {Booth}, R., {Hafez}, Y., \& {Almeqren}, E. 2010, in Bulletin of the American
  Astronomical Society, vol.~41 of Bulletin of the American Astronomical
  Society, 602

\bibitem[{{Page} et~al.(2007){Page}, {Hinshaw}, {Komatsu}, {Nolta}, {Spergel},
  {Bennett}, {Barnes}, {Bean}, {Dor{\'e}}, {Dunkley}, {Halpern}, {Hill},
  {Jarosik}, {Kogut}, {Limon}, {Meyer}, {Odegard}, {Peiris}, {Tucker}, {Verde},
  {Weiland}, {Wollack}, \& {Wright}}]{page07}
{Page}, L., {Hinshaw}, G., {Komatsu}, E., {Nolta}, M.~R., {Spergel}, D.~N.,
  {Bennett}, C.~L., {Barnes}, C., {Bean}, R., {Dor{\'e}}, O., {Dunkley}, J.,
  {Halpern}, M., {Hill}, R.~S., {Jarosik}, N., {Kogut}, A., {Limon}, M.,
  {Meyer}, S.~S., {Odegard}, N., {Peiris}, H.~V., {Tucker}, G.~S., {Verde}, L.,
  {Weiland}, J.~L., {Wollack}, E., \& {Wright}, E.~L. 2007, \apjs, 170, 335.
  \eprint{arXiv:astro-ph/0603450}

\bibitem[{{Samtleben} \& {for the QUIET collaboration}(2008)}]{samtleben08}
{Samtleben}, D., \& {for the QUIET collaboration} 2008, ArXiv e-prints.
  \eprint{0806.4334}

\bibitem[{{Testori} et~al.(2008){Testori}, {Reich}, \& {Reich}}]{testori08}
{Testori}, J.~C., {Reich}, P., \& {Reich}, W. 2008, \aap, 484, 733

\bibitem[{{Tucci} et~al.(2005){Tucci}, {Mart{\'{\i}}nez-Gonz{\'a}lez},
  {Vielva}, \& {Delabrouille}}]{tucci05}
{Tucci}, M., {Mart{\'{\i}}nez-Gonz{\'a}lez}, E., {Vielva}, P., \&
  {Delabrouille}, J. 2005, \mnras, 360, 935. \eprint{arXiv:astro-ph/0411567}

\bibitem[{{Wolleben} et~al.(2006){Wolleben}, {Landecker}, {Reich}, \&
  {Wielebinski}}]{wolleben06}
{Wolleben}, M., {Landecker}, T.~L., {Reich}, W., \& {Wielebinski}, R. 2006,
  \aap, 448, 411. \eprint{arXiv:astro-ph/0510456}

\bibitem[{{Zaldarriaga}(1998)}]{zaldarriaga98}
{Zaldarriaga}, M. 1998, Ph.D. thesis, MASSACHUSETTS INSTITUTE OF TECHNOLOGY

\end{thebibliography}

\end{document}